\documentclass{article}
\usepackage{amsfonts}
\usepackage{amsmath}
\usepackage{graphicx}
\usepackage{epsfig}

\begin{document}

\title{Extending canonical Monte Carlo methods II}
\author{L. Velazquez and S. Curilef \\
Departamento de F\'{\i}sica, Universidad Cat\'{o}lica del Norte, \\
Av. Angamos 0610, Antofagasta, Chile.\\
E-mail: lvelazquez@ucn.cl and scurilef@ucn.cl}
\maketitle
\tableofcontents

\begin{abstract}
Previously, we have presented a methodology to extend canonical
Monte Carlo methods inspired on a suitable extension of the
canonical fluctuation relation $C=\beta^{2}\left\langle\delta
E^{2}\right\rangle$ compatible with negative heat capacities $C<0$.
Now, we improve this methodology by introducing a better treatment
of finite size effects affecting the precision of a direct
determination of the microcanonical
caloric curve $\beta \left( E\right) =\partial S\left( E\right) /\partial E$%
, as well as a better implementation of MC schemes. We shall show
that despite the modifications considered, the extended canonical MC
methods possibility an impressive overcome of the so-called \textit{%
super-critical slowing down} observed close to the region of a
temperature driven first-order phase transition. In this case, the
dependence of the decorrelation time $\tau$ with the system size $N$
is reduced from an exponential growth to a weak power-law behavior
$\tau(N)\propto N^{\alpha}$, which is shown in the particular case
of the 2D seven-state Potts model where the exponent
$\alpha=0.14-0.18$.
\end{abstract}

\section{Introduction}

In a previous paper \cite{vel-emc}, we have proposed a methodology that
enables Monte Carlo (MC)\ methods based on the Gibbs canonical ensemble:
\begin{equation}
dp_{c}\left( E\left\vert \beta _{B}\right. \right) =\frac{1}{Z\left( \beta
_{B}\right) }\exp \left( -\beta _{B}E\right) \Omega \left( E\right) dE
\label{can}
\end{equation}%
to account for the existence of an anomalous region with negative heat
capacities $C<0$ \cite{pad,Lyn3,gro1,gro na,moretto,Dagostino} and to
overcome the so-called \textit{supercritical slowing down} observed near of
first-order phase transition \cite{mc3}. This development is inspired on the
consideration of a recently obtained fluctuation relation \cite%
{vel-jpa,vel-jstat,vel-geo}:
\begin{equation}
C=\beta ^{2}\left\langle \delta E^{2}\right\rangle +C\left\langle \delta
\beta _{\omega }\delta E\right\rangle ,  \label{fdr}
\end{equation}%
which appears as a suitable extension of the known canonical
identity $ C=\beta ^{2}\left\langle \delta E^{2}\right\rangle $
involving the heat capacity $C$ and the energy fluctuations
$\left\langle \delta E^{2}\right\rangle $. This last expression
accounts for the realistic perturbation provoked on the internal
state of a certain environment as a consequence of the underlying
thermodynamic interaction with the system under study, which is
characterized here in terms of the correlation function
$\left\langle \delta \beta _{\omega }\delta E\right\rangle $ between
the system internal energy $E$ and the environment inverse
temperature $ \beta _{\omega }$. While the canonical ensemble
(\ref{can}) dismisses the existence of this feedback effect (due to
the constancy of the inverse temperature $\beta _{B}$ associated
with this ensemble) and it is only compatible with positive heat
capacities, the general case possibilities to access anomalous
macrostates with negative heat capacities $C<0$ as long as the
condition $ \left\langle \delta \beta _{\omega }\delta
E\right\rangle >1$ is obeyed.

The incidence of non-vanishing correlated fluctuations $\left\langle \delta
\beta _{\omega }\delta E\right\rangle$ can be easily implemented in MC
simulations. Roughly speaking, the extension of canonical MC methods is
achieved by replacing the canonical inverse temperature $\beta _{B}$ with a
\textit{variable inverse temperature} $\beta _{\omega }\left( E\right) $.
The resulting framework constitutes a suitable extension of the Gerling and H%
\"{u}ller methodology on the basis of the so-called \textit{dynamic
ensemble} \cite{GerlingHuller}, where the microcanonical curve
$\beta \left(
E\right) =\partial S\left( E\right) /\partial E$ and the heat capacity $%
C\left( E\right) $ can be estimated in terms of the energy and
temperature expectation values $\left\langle E\right\rangle $ and
$\left\langle \beta _{\omega }\right\rangle $ as well as their
fluctuating behavior described by Eq.(\ref{fdr}). This method
successfully reduces the exponential divergence of decorrelation
time $\tau \propto \exp \left( \lambda N\right) $ with the increase
of the system size $N$ of canonical MC methods to a weak power-law
divergence $\tau \propto N^{\alpha }$, with a typical exponent
$\alpha \simeq 0$.$2$ for the case of $2D$ ten-state Potts model
\cite{vel-emc}. By combining this type of argument with cluster
algorithms, one obtains very efficient MC schemes that constitute
attractive alternatives of the known multicanonical method and its
variants \cite{mc3}. In this work, we shall improve the present
methodology to consider the existence of finite size effects that
reduces the precision of a direct determination of the
microcanonical caloric curve $\beta \left( E\right) =\partial
S\left( E\right) /\partial E$, as well as a better implementation of
MC schemes.

\section{Methodology}

\subsection{Overview}

The simplest way to implement the existence of non-vanishing
correlations $ \left\langle \delta \beta _{\omega }\delta
E\right\rangle$ corresponds to a linear coupling of the environment
inverse temperature $\beta _{\omega }$ with the thermal fluctuations
of the system energy:
\begin{equation}
\beta _{\omega }\left( E\right) =\beta _{e}+\lambda \delta E/N,
\label{linear.ansatz}
\end{equation}%
where $\lambda $ is a coupling constant that appears here as an
additional control parameter. While the case with $\lambda =0$
corresponds to the canonical ensemble (\ref{can}), where $\beta
_{\omega} =const$, in general, the constancy of the inverse
temperature $\beta _{e}$ can be only ensured in the average sense,
$\beta _{e}=\left\langle \beta _{\omega }\right\rangle $.
Eq.(\ref{linear.ansatz}) can be substituted into Eq.(\ref{fdr}) to
obtain the following results:
\begin{equation}
\left( \Delta E\right) ^{2}=\frac{N}{\beta ^{2}N/C+\lambda },~\left( \Delta
\beta _{\omega }\right) ^{2}=\frac{1}{N}\frac{\lambda ^{2}}{\beta
^{2}N/C+\lambda },  \label{dispersions}
\end{equation}%
where $\Delta x\equiv \sqrt{\left\langle \delta x^{2}\right\rangle }$
denotes the thermal dispersion of a given quantity $x$. Since $\Delta \beta
_{\omega }$ and $\Delta E$ should be nonnegative, one arrives at the
following stability condition:
\begin{equation}
\beta ^{2}N/C+\lambda >0.  \label{stab.cond}
\end{equation}%
For $\lambda =0$, one derives the ordinary constraint $C>0$ that
emphasizes the unstable character of macrostates with negative heat
capacities $C<0$ within the canonical description. However, these
anomalous
macrostates can be observed in a stable way in a general situation with $%
\lambda \not=0$ when this control parameter satisfies Eq.(\ref%
{stab.cond}). By assuming an extensive character of the heat
capacity $C\sim N$ in short-range interacting systems, the energy
dispersion $\Delta E$ grows with the increase of the system size $N$
as $\Delta E\sim \sqrt{N}$, so that, the dispersion of the energy
per particle $\varepsilon =E/N$ behaves as $\Delta \varepsilon \sim 1/\sqrt{N%
}$. Since $\delta \beta _{\omega }\equiv \lambda \delta E/N$ in the
case of the ansatz (\ref{linear.ansatz}), the dispersion of the
inverse temperature also behaves as $\Delta \beta _{\omega }\sim
1/\sqrt{N}$. Thus, the present equilibrium situation constitutes a
physical scenario that ensures the stability of macrostates with
negative heat capacities $C<0$ with the incidence of small thermal
fluctuations.

\begin{figure}[t]
\begin{center}
\includegraphics[
height=2.2373in, width=3.0381in
]{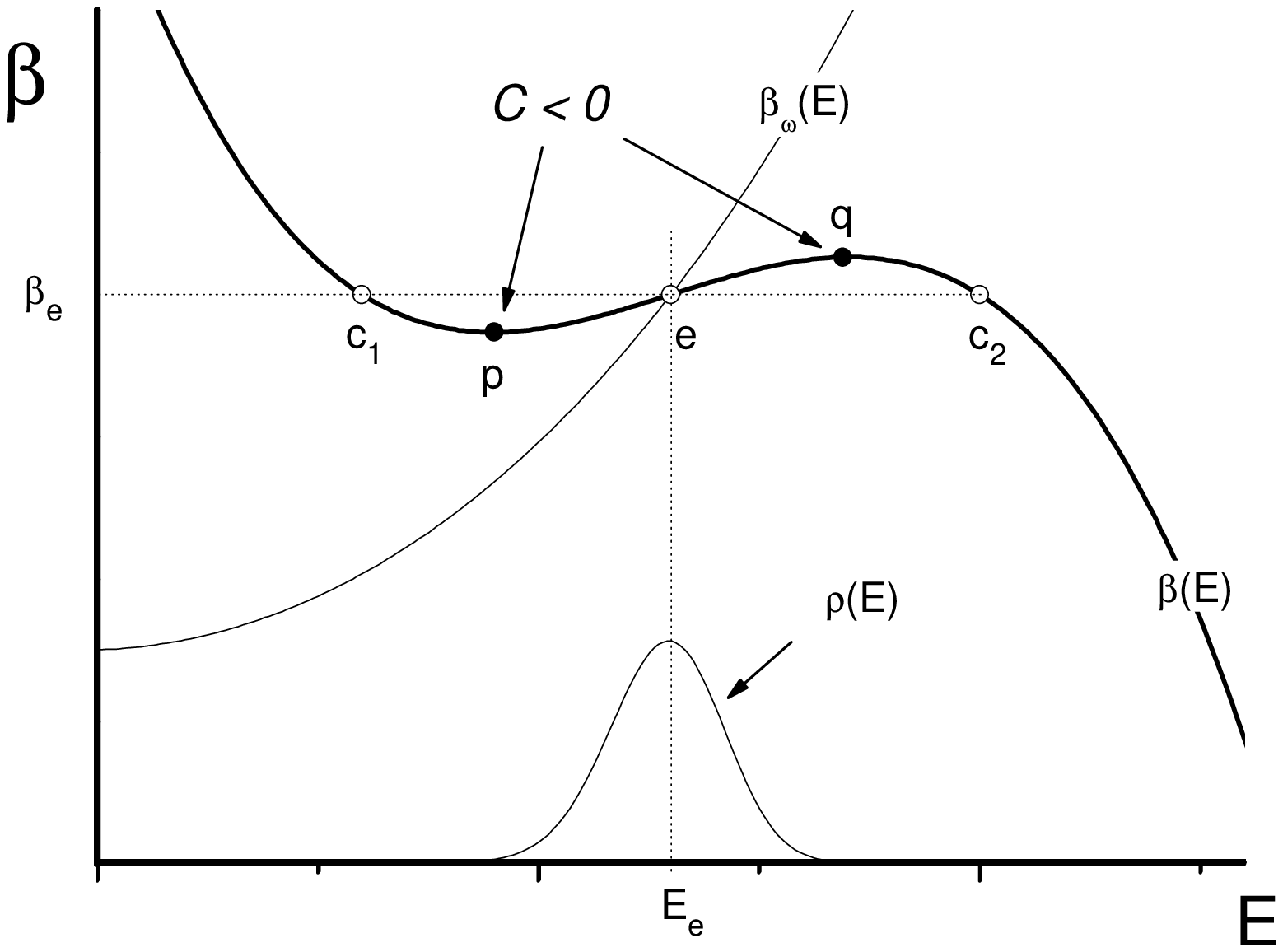}
\end{center}
\caption{Schematic representation of the typical microcanonical caloric
curve $\protect\beta\left( E\right) =\partial S\left( E\right) /\partial E$
associated with a finite short-range interacting system undergoing a
first-order phase transition, as well as the energy distribution function $%
\protect\rho\left( E\right) $ resulting from the thermal coupling of this
system with a certain environment with inverse temperature $\protect\beta_{%
\protect\omega}\left( E\right) $. For further explanations, see the text.}
\label{caricature.eps}
\end{figure}

This type of equilibrium situation is schematically represented in FIG.\ref%
{caricature.eps}. Here, it is shown the typical microcanonical caloric curve $%
\beta \left( E\right) =\partial S\left( E\right) /\partial E$
corresponding to a finite short-range interacting system that
undergoes a first-order phase transition, as well as the energy
distribution function $\rho \left( E\right) $ associated with the
thermal coupling of this system with an environment with inverse
temperature $\beta _{\omega }\left( E\right) $. The intersection
points $E_{e}$ derived from the condition of thermal equilibrium:
\begin{equation}  \label{cond.thermal}
\beta \left( E_{e}\right) =\beta _{\omega }\left( E_{e}\right)
\end{equation}%
determine the position of maxima and minima of the distribution function $%
\rho \left( E\right) $. Note that the linear ansatz
(\ref{linear.ansatz}) can be considered as the first-order
approximation of the power expansion of a general dependence $
\beta_{\omega}(E)$:
\begin{equation}
\beta_{\omega}(E)=\beta_{e}+\sum^{\infty}_{n=1}a_{n}(E-E_{e})^{n},
\end{equation}
where $a_{1}=\partial\beta_{\omega}(E_{e})/\partial
E\equiv\lambda/N$. While in the canonical ensemble there could exist
three intersections points ($e$, $c_{1}$ and $c_{2}$) because of the
constancy of the inverse temperature, it is possible to ensure the
existence of only one intersection point $E_{e}$ by appropriately
choosing the inverse temperature dependence $\beta _{\omega }\left(
E\right) $. If the system size $N$ is sufficiently large, the energy
distribution function\ $\rho \left( E\right) $ adopts a bell shape,
which is approximately described with
the Gaussian distribution:%
\begin{equation}
\rho \left( E\right) \simeq A\exp \left[ -\frac{1}{2\sigma _{E}^{2}}\left(
E-E_{e}\right) ^{2}\right] ,  \label{GD}
\end{equation}%
where $\sigma _{E}=\Delta E$. The expectation values $\left\langle
E\right\rangle $ and $\left\langle \beta _{\omega }\right\rangle $ provide a
direct estimation of the energy $E_{e}$ and inverse temperature $\beta
_{e}=\beta \left( E_{e}\right)$ at the intersection point illustrated in FIG.%
\ref{caricature.eps}:
\begin{equation}
E_{e}=\left\langle E\right\rangle ,~\beta _{e}=\left\langle \beta _{\omega
}\right\rangle .  \label{work1}
\end{equation}

The procedure previously described is equivalent to the one employed
by Gerling and H\"{u}ller in the framework of the dynamical ensemble
\cite{GerlingHuller}:
\begin{equation}
\omega _{D}\left( E\right) =A\left( E_{T}-E\right) ^{B},  \label{dyn}
\end{equation}%
which accounts for the thermal coupling of the system with a bath exhibiting
a constant heat capacity, e.g., an ideal gas, whose inverse temperature
obeys the following dependence on the system energy $E$:
\begin{equation}
\beta _{\omega }\left( E\right) =B/\left( E_{T}-E\right) .  \label{GHb}
\end{equation}%
Our proposal considers some improvements to the Gerling and Huller's
methodology. In fact, the energy dependence (\ref{GHb}) is less convenient
for calculations than the linear ansatz (\ref{linear.ansatz}), which is a
feature particularly useful to perform the analysis of \textit{finite size
effects }(see subsection \ref{finite} below). Moreover, one can also obtain
the value of the heat capacity $C\left( E\right) $, or more exactly, the
so-called \textit{curvature curve} $\kappa \left( E\right) $:%
\begin{equation}
\kappa \left( E\right) =\beta ^{2}N/C\equiv -N\frac{\partial ^{2}S\left(
E\right) }{\partial E^{2}}  \label{curvatura}
\end{equation}%
at the intersection point $E_{e}$, $\kappa _{e}=\kappa \left( E_{e}\right) $%
, through the energy dispersion $\Delta E$ in analogous way of MC
calculation by using the canonical ensemble (\ref{can}):
\begin{equation}
\kappa _{e}=\frac{1-\lambda \left( \Delta E\right) ^{2}/N}{\left( \Delta
E\right) ^{2}/N}.  \label{curvature.fluc}
\end{equation}%
This last equation was obtained from Eq.(\ref{curvatura}) by rewriting the
first relation of Eq.(\ref{dispersions}).

Since the energy and the inverse temperature dispersions in Eq.(\ref%
{dispersions}) are controlled by the coupling constant $\lambda $, it is
desirable to reduce them as low as possible. One can verify that the energy
dispersion $\Delta E$ decreases with the increase of the coupling constant $%
\lambda $. However, the value of this parameter should not be excessively
large. Its increment leads to the increase of the inverse temperature
dispersion $\Delta \beta _{\omega }$, which affects the precision of the
inverse temperature $\beta _{e}$ of the system indirectly derived from the
expectation value $\left\langle \beta _{\omega }\right\rangle $. A
thermodynamic criterion to provide an optimal value for the coupling
constant $\lambda $ can be obtained by minimizing the total dispersion $%
\Delta _{T}^{2}$:
\begin{equation}
\Delta _{T}^{2}=\frac{\left( \Delta E\right) ^{2}}{N}+N\left( \Delta \beta
_{\omega }\right) ^{2}=\frac{1+\lambda ^{2}}{\lambda +\kappa _{e}},
\label{min.DeltaT}
\end{equation}%
which talks about the precision in the determination of the intersection
point $\left( E_{e},\beta _{e}\right) $. This analysis leads to the
following result:
\begin{equation}
\lambda _{\Delta }=\lambda _{\Delta }\left( \kappa _{e}\right) =\sqrt{%
1+\kappa _{e}^{2}}-\kappa _{e},~\min \Delta _{T}^{2}=2\lambda _{\Delta }
\label{optimal}
\end{equation}

\subsection{Finite size effects\label{finite}}

The energy distribution function corresponding to the thermal
coupling of the system with an environment can be expressed by the
following equation:
\begin{equation}
\rho \left( E\right) dE=\omega \left( E\right) \Omega \left( E\right) dE.
\label{dist}
\end{equation}%
Here, $\omega \left( E\right) $ is a probabilistic weight that
characterizes such a thermodynamic influence, which is related to
the environment inverse temperature $\beta _{\omega }\left( E\right)
$ as follows:
\begin{equation}
\beta _{\omega }\left( E\right) =-\frac{\partial \log \omega \left( E\right)
}{\partial E}.
\end{equation}%
A direct integration allows us to verify that the probabilistic weight
associated with the linear ansatz (\ref{linear.ansatz}) is simply the
\textit{Gaussian ensemble} \cite{Challa}:
\begin{equation}
\omega _{G}\left( E\right) =\frac{1}{Z_{\lambda }\left( \beta _{e}\right) }%
\exp \left[ -\beta _{e}E-\frac{1}{2N}\lambda \left( E-E_{e}\right) ^{2}%
\right]  \label{GEns}
\end{equation}%
introduced by Hetherington \cite{Hetherington}, which approaches in the
limit $\lambda \rightarrow +\infty $ to the microcanonical ensemble:%
\begin{equation}
\omega _{M}=\frac{1}{\Omega }\delta \left( E-E_{e}\right) .
\end{equation}

As already explained, the estimation of the microcanonical caloric curve $%
\beta \left( E\right) $, as well as the fluctuation relation (\ref{fdr}),
are based on the consideration of a Gaussian shape for the energy
distribution function $\rho \left( E\right) $. Such an approximation
naturally arises as the asymptotic distribution as long as the system size $%
N $ be sufficiently large. If the system size $N$ is not so large, small
deviations from the Gaussian profile (\ref{GD}) are naturally expected.
Fortunately, the particular mathematical form of the Gaussian ensemble (\ref%
{GEns}) possibilities to consider some simple corrections formulae to deal
with the existence of these finite size effects and to improve the precision
of the present methodology.

Let us denote the energy distribution function in terms of the energy per
particle $\varepsilon =E/N$ as follows:
\begin{equation}
\rho \left( \varepsilon \right) =\frac{1}{Z_{\lambda }}\exp \left[ -N\phi
\left( \varepsilon \right) \right] ,
\end{equation}%
where the function $\phi \left( \varepsilon \right) \equiv \beta
_{e}\varepsilon +\frac{1}{2}\lambda \left( \varepsilon -\varepsilon
_{e}\right) ^{2}-s\left( \varepsilon \right) $, with $s\left( \varepsilon
\right) $ being the entropy per particle, and $\beta _{e}$:%
\begin{equation}
\beta _{e}=\frac{\partial s\left( \varepsilon _{e}\right) }{\partial
\varepsilon },
\end{equation}%
the system inverse temperature at the stationary point $\varepsilon _{e}$.
Let us now develop a power-expansion:
\begin{equation}
\phi \left( \varepsilon _{e}+x\right) =\phi \left( \varepsilon _{e}\right)+%
\frac{1}{2}\left( \lambda +\kappa _{e}\right) x^{2}+\sum_{n=3}^{\infty
}a_{n}x^{n},  \label{power}
\end{equation}%
with $\kappa _{e}=-\partial ^{2}s\left( \varepsilon _{e}\right) /\partial
\varepsilon ^{2}$ being the curvature. The Gaussian approximation is
developed by dismissing those terms with $n>2$. Here, the energy deviation $%
x $ obeys a Gaussian distribution:%
\begin{equation}
\rho ^{\left( 0\right) }\left( x\right) dx\simeq \frac{1}{\sqrt{2\pi
}\sigma }\exp \left( -\frac{x^{2}}{2\sigma ^{2}}\right) dx
\label{Gaussian}
\end{equation}%
with standard deviation $\sigma $:
\begin{equation}
\sigma ^{2}=\frac{1}{N\left( \lambda +\kappa _{e}\right) },  \label{sd}
\end{equation}%
and the partition function $Z_{\lambda }$ can be approximated as:
\begin{equation}
Z_{\lambda }=e^{-N\phi(\varepsilon_{e})}\sqrt{2\pi }\sigma .
\label{Z.lambda}
\end{equation}%
Note that Eq.(\ref{sd}) is fully equivalent to Eq.(\ref{curvature.fluc})
since the standard deviation $\sigma\equiv\Delta E/N$. Eq.(\ref{Z.lambda})
can be rewritten as follows:
\begin{equation}
P_{\lambda }\simeq \beta _{e}E_{e}-S\left( E_{e}\right) +\frac{1}{2}\log
\left( 2\pi \sigma ^{2}\right) ,
\end{equation}%
where $P_{\lambda }=-\log Z_{\lambda }$ can be referred to as the Plank
thermodynamic potential corresponding to the Gaussian ensemble (\ref{GEns}).
By performing the thermodynamic limit $N\rightarrow \infty $, this
thermodynamic function can be related to the known Legendre transformation
with the microcanonical entropy:
\begin{equation}
p_{\lambda }=\lim_{N\rightarrow \infty }\frac{P_{\lambda }}{N}=\beta
_{e}\varepsilon _{e}-s\left( \varepsilon _{e}\right) .
\end{equation}%
Clearly, the Gaussian ensemble (\ref{GEns}) provides a suitable extension of
the Gibbs canonical ensemble (\ref{can}) that is able to deal with the
existence of macrostates with negative heat capacities $C<0$. In fact, it is
a particular example of the so-called \textit{generalized canonical ensembles%
} that preserve some of its more relevant features \cite%
{vel-geo,hugo.gce,toral}.

The first correction of the Gaussian approximation (\ref{Gaussian}) is
obtained by dismissing terms with $n>3$ in the power expansion (\ref{power}%
):
\begin{equation}
N\phi \left( x\right) =N\phi _{0}+\frac{1}{2\sigma ^{2}}x^{2}+\xi
x^{3}+O\left( x^{4}\right) ,
\end{equation}%
where:
\begin{equation}
\xi =-\frac{1}{6}N\frac{\partial ^{3}s\left( \varepsilon _{e}\right) }{%
\partial \varepsilon ^{3}}.
\end{equation}%
It is convenient to introduce the dimensionless variable $\theta=x/\sigma $.
By considering the size dependencies $\xi \sim N$ and $\sigma \sim 1/\sqrt{N}
$, it is possible to verify that the cubic term $\xi x^{3}\equiv \xi \sigma
^{3}\theta ^{3}$ decreases as $1/ \sqrt{N}$ with the increase of the system
size $N$. Thus, one arrives at the distribution function:
\begin{equation}
\rho ^{\left( 1\right) }\left( \theta \right) d\theta \simeq \frac{1}{\sqrt{%
2\pi }}e^{-\frac{1}{2}\theta ^{2}}\left( 1-\xi \sigma ^{3}\theta ^{3}\right)
d\theta .
\end{equation}%
This last result leads to the following expectation values:
\begin{equation}
\left\langle \theta \right\rangle =-3\xi \sigma ^{3},~\left\langle \theta
^{2}\right\rangle =1,~\left\langle \theta ^{3}\right\rangle =-15\xi \sigma
^{3},
\end{equation}%
which allow to express the expectation value of the energy deviation $%
\left\langle x\right\rangle $ as well as its second and third order
dispersions, $\left\langle \delta x^{2}\right\rangle $ and $\left\langle
\delta x^{3}\right\rangle $, as follows:
\begin{eqnarray}
\left\langle x\right\rangle &=&-3\xi \sigma ^{4}+O\left( \frac{1}{N^{2}}%
\right),\left\langle \delta x^{2}\right\rangle =\sigma ^{2}+O\left( \frac{1}{%
N^{2}}\right) ,  \label{mean} \\
~\left\langle \delta x^{3}\right\rangle &=&-6\xi \sigma ^{6}+O\left( \frac{1%
}{N^{3}}\right) .  \label{fluct1}
\end{eqnarray}%
The previous results can be combined with the linear ansatz (\ref%
{linear.ansatz}) to obtain the following first-order correction of Eq.(\ref%
{work1}):
\begin{align}
E_{e}& =\left\langle E\right\rangle -\frac{1}{2\left\langle \delta
E^{2}\right\rangle }\left\langle \delta E^{3}\right\rangle ,  \label{e1} \\
\beta _{e}& =\left\langle \beta _{\omega }\right\rangle -\lambda \frac{1}{%
2N\left\langle \delta E^{2}\right\rangle }\left\langle \delta
E^{3}\right\rangle .  \label{e2}
\end{align}

The second-order correction of the Gaussian constributions is carried out by
dismissing terms with $n>4$ in power expansion (\ref{power}):
\begin{equation}
N\phi\left( x\right) =\frac{1}{2\sigma^{2}}x^{2}+\xi
x^{3}+\xi_{2}x^{4}+O\left( x^{5}\right) ,
\end{equation}
where:
\begin{equation}
\xi_{2}=-\frac{1}{24}N\frac{\partial^{4}s\left( \varepsilon_{e}\right) }{%
\partial\varepsilon^{4}}.
\end{equation}
These terms lead to the following correction of the distribution function:
\begin{equation}
\rho^{\left( 2\right) }\left( \theta\right) d\theta\simeq\frac {e^{-\frac{1}{%
2}\theta^{2}}}{A\sqrt{2\pi}}\left( 1-\xi\sigma^{3}\theta
^{3}-\xi_{2}\sigma^{4}\theta^{4}+\frac{1}{2}\xi^{2}\sigma^{6}\theta
^{6}\right) d\theta.
\end{equation}
whose third and fourth terms account for finite size effects of order $%
O\left( 1/N\right) $. By denoting the auxiliary constants $C_{1}$ and $C_{2}$
as follows:%
\begin{equation}
C_{1}=\xi\sigma^{3},~C_{2}=\xi_{2}\sigma^{4},
\end{equation}
direct calculations allow to obtain the normalization constant $A$:
\begin{equation}
A=1-3C_{2}+\frac{15}{2}C_{1}^{2},
\end{equation}
as well as the following expectation values:
\begin{align}
\left\langle \theta\right\rangle & =-3C_{1},~\left\langle \theta
^{2}\right\rangle =1-12C_{2}+45C_{1}^{2}, \\
\left\langle \theta^{3}\right\rangle & =-15C_{1},~\left\langle \theta
^{4}\right\rangle =3-102C_{2}+465C_{1}^{2}.
\end{align}
While the expressions (\ref{mean}) and (\ref{fluct1}) for the expectation
values $\left\langle x\right\rangle $ and $\left\langle \delta
x^{3}\right\rangle $ remain invariable under the second-order approximation,
the second and fourth order dispersions $\left\langle \delta
x^{2}\right\rangle $ and $\left\langle \delta x^{4}\right\rangle $ exhibit
the following corrections:
\begin{align}
\left\langle \delta x^{2}\right\rangle & =\sigma^{2}\left(
1-12C_{2}+36C_{1}^{2}\right) +O\left( \frac{1}{N^{3}}\right) , \\
\left\langle \delta x^{4}\right\rangle & =\left(
3-102C_{2}+339C_{1}^{2}\right) \sigma^{4}+O\left( \frac{1}{N^{4}}\right) .
\end{align}
By introducing the cumulants $\epsilon_{1}$ and $\epsilon_{2}$:%
\begin{equation}
\epsilon_{1}=\frac{\left\langle \delta E^{3}\right\rangle ^{2}}{\left\langle
\delta E^{2}\right\rangle ^{3}},~\epsilon_{2}=1-\frac{\left\langle \delta
E^{4}\right\rangle }{3\left\langle \delta E^{2}\right\rangle ^{2}},
\end{equation}
the auxiliary constants $C_{1}^{2}$ and $C_{2}$ can be expressed as follows:%
\begin{equation}
C_{1}^{2}=\frac{1}{36}\epsilon_{1},~C_{2}=\frac{1}{10}\epsilon_{2}+\frac {41%
}{360}\epsilon_{1}
\end{equation}
Thus, the main work equations (\ref{work1}) and (\ref{curvature.fluc}) can
be expressed in this second-order approximation as follows:
\begin{align}
E_{e} & =\left\langle E\right\rangle -\frac{1-\psi_{1}}{2\left\langle \delta
E^{2}\right\rangle }\left\langle \delta E^{3}\right\rangle ,  \label{EE} \\
\beta_{e} & =\left\langle \beta_{\omega}\right\rangle -\lambda\frac {%
1-\psi_{1}}{2N\left\langle \delta E^{2}\right\rangle }\left\langle \delta
E^{3}\right\rangle , \\
\kappa_{e} & =\frac{1-\psi_{1}-\lambda\left( \Delta E\right) ^{2}/N}{\left(
\Delta E\right) ^{2}/N},  \label{EC}
\end{align}
where $\psi_{1}$ is a second-order term defined by the cumulants $%
\epsilon_{1}$ and $\epsilon_{2}$ as:
\begin{equation}
\psi_{1}\equiv\frac{6}{5}\epsilon_{2}+\frac{11}{30}\epsilon_{1}.
\end{equation}
It is easy to verify that the surviving finite size effects of the above
correction for the caloric curve $\beta$ in terms of the energy per particle
$\varepsilon$ are of order $O\left( 1/N^{3}\right) $, while the ones
corresponding to the curvature curve $\kappa$ \textit{versus} $\varepsilon$
are of order $O\left( 1/N^{2}\right) $.

As a by-product of the previous analysis, one can obtain the third and
fourth derivatives of the entropy per particle $s\left( \varepsilon\right) $
as follows:
\begin{align}
\zeta _{e}^{3}& =\frac{\partial ^{3}s\left( \varepsilon _{e}\right) }{%
\partial \varepsilon ^{3}}=N^{2}\frac{\left\langle \delta E^{3}\right\rangle
}{\left\langle \delta E^{2}\right\rangle ^{3}}\left( 1-3\psi _{1}\right) ,
\label{z3} \\
\zeta _{e}^{4}& =\frac{\partial ^{4}s\left( \varepsilon _{e}\right) }{%
\partial \varepsilon ^{4}}=-\psi _{2}\frac{N^{3}}{\left\langle \delta
E^{2}\right\rangle ^{2}},  \label{z4}
\end{align}%
where $\psi _{2}$ is another second-order term defined by cumulants $%
\epsilon _{1}$ and $\epsilon _{2}$ as:
\begin{equation}
\psi _{2}=\frac{12}{5}\epsilon _{2}+\frac{41}{15}\epsilon _{1}.  \label{z5}
\end{equation}%
The underlying finite size effects of the dependence $\zeta _{3}$ \textit{%
versus} $\varepsilon $ estimated by Eq.(\ref{z3}) are of order $O\left(
1/N^{2}\right) $, while the ones corresponding to the dependence $\xi _{4}$
\textit{versus} $\varepsilon $ estimated by Eq.(\ref{z4}) are of order $%
O\left( 1/N\right) $.

\subsection{Implementation}

As already commented in the introductory section, a general way to extend a
particular canonical MC algorithm with transition probability $W\left(
X_{i}\rightarrow X_{f};\beta _{B}\right) $ by using the present methodology
is achieved by replacing the inverse temperature $\beta _{B}$ of the
canonical ensemble (\ref{can}) with a variable inverse temperature, $\beta
_{B}\rightarrow \beta _{\omega }\left( E\right) $. The fulfilment of the
detailed balance condition:
\begin{equation}
p_{\omega }\left( X_{i}\right) W\left( X_{i}\rightarrow X_{j};\beta _{\omega
}^{t}\right) =p_{\omega }\left( X_{j}\right) W\left( X_{j}\rightarrow
X_{i};\beta _{\omega }^{t}\right)  \label{detailed.balance}
\end{equation}%
demands to use a certain value $\beta _{\omega }^{t}$\ of the
environment inverse temperature $\beta _{B}$ for both the direct and
reverse process defined from the condition:
\begin{equation}
\frac{p_{\omega }\left( X_{i}\right) }{p_{\omega }\left( X_{f}\right) }=\exp
\left( \beta _{\omega }^{t}\delta E_{if}\right),
\end{equation}
which is hereafter referred to as the \textit{transition inverse temperature}%
. Here, $p_{\omega }\left( X\right) \equiv \omega \left[ E\left( X\right) %
\right] $ represents the distribution function associated with the
environment inverse temperature $\beta _{\omega }\left( E\right) $,
and $\delta E_{if}=E_{j}-E_{i}$, the energy varying of the system
during the transition, where $E_{i}=E\left( X_{i}\right) $ and
$E_{f}=E\left( X_{f}\right) $. The case of the linear ansatz
(\ref{linear.ansatz}) is special, since its transition inverse
temperature $\beta _{\omega }^{t}$ is simply given by:
\begin{equation}
\beta _{\omega }^{t}=\frac{1}{2}\left( \beta _{\omega }^{i}+\beta _{\omega
}^{f}\right) ,  \label{eq.temp}
\end{equation}%
where $\beta _{\omega }^{i}$ and $\beta _{\omega }^{f}$ are the bath inverse
temperatures at the initial and the final configurations respectively, $%
\beta _{\omega }^{i}=\beta _{\omega }\left( E_{i}\right) $ and $\beta
_{\omega }^{f}=\beta _{\omega }\left( E_{f}\right) $.

The direct applicability of the above result is restricted due to the final
configuration $X_{f}$ must be previously known in order to obtain the exact
value of the transition inverse temperature $\beta _{\omega }^{t}$ . While
such a requirement can be always satisfied in a local MC such as Metropolis
importance sample \cite{metro,Hastings}, the final configuration $X_{f}$ is
\textit{a priori} unknown in non-local MC methods such as clusters
algorithms \cite%
{SW,pottsm,wolf,Edwards,Niedermayer,Evertz,Hasenbusch,Dress,Liu}.
For these cases, one is forced to employ an approximated value of
the transition inverse temperature $\beta _{\omega }^{t}$, e.g.: the
inverse temperature $\beta _{\omega }^{i}$ at the initial
configuration, $\beta _{\omega }^{i}=\beta _{\omega }\left(
E_{i}\right) $. Although the resulting MC algorithm does not obey
the detailed balance, we have shown that the deviation of the
asymptotic distribution function $\tilde{p}_{\omega }\left( X\right)
$ from the exact distribution function $p_{\omega }\left( X\right) $
can be disregarded for $N$ sufficiently large. This is the reason
why this method is particularly useful to overcome slow sampling
problems in large scale MC simulations.

The use of an approximated value for the transition inverse temperature $%
\beta _{\omega }^{t}$ is not longer appropriated when one is also interested
in the study of systems with size $N$ relatively small. Such an
approximation introduces uncontrollable finite size effects that cannot be
dealt by using work equations (\ref{EE}-\ref{EC}). In this case, it is
necessary to fulfil the detailed balance condition (\ref{detailed.balance})
to obtain the Gaussian profile (\ref{GEns}) as asymptotic distribution
function. The most general way to achieve this aim is to introduce \textit{a
posteriori} acceptance probability $w_{i\rightarrow f}$:
\begin{equation}
w_{i\rightarrow f}=\min \left\{ 1,\frac{W_{f\rightarrow i}}{W_{i\rightarrow
f}}\exp \left( -\beta _{\omega }^{t}\delta E_{if}\right) \right\}
\label{wif}
\end{equation}%
to accept or reject the final configuration $X_{f}$. Here, the terms:
\begin{equation}
W_{i\rightarrow f}=W\left[ X_{i}\rightarrow X_{f};\beta _{\omega }^{i}\right]%
,W_{f\rightarrow i}=W\left[ X_{f}\rightarrow X_{i};\beta _{\omega }^{f}%
\right]
\end{equation}
represent the transition probabilities of the direct and the reverse
process, respectively, which should be calculated for the given canonical
clusters MC algorithm.

\subsection{Iterative schemes}

Given a certain dependence of the environment inverse temperature
$\beta _{\omega }^{\left( i\right) }\left( E\right) $, one can
obtain from a MC run a punctual estimation of the system inverse
temperature $\beta _{i}$, the curvature $\kappa _{i}$, as well as
the third and fourth derivatives of the entropy per particle $\zeta
_{i}^{3}$ and $\zeta _{i}^{4}$ at the $i$-th intersection point
$\varepsilon _{i}$. These values can be considered to provide the
next dependence $\beta _{\omega }^{\left( i+1\right) }\left(
E\right) $. The linear ansatz (\ref{linear.ansatz}) can be rewritten
in terms of the energy per particle as follows:
\begin{equation}
\beta _{\omega }=\beta _{i}^{\ast }+\lambda _{i}\left( \varepsilon
-\varepsilon _{i}^{\ast }\right) ,  \label{seeds}
\end{equation}%
where $\varepsilon _{i}^{\ast }$ and $\beta _{i}^{\ast }$ are roughly
estimations of the correct values $\varepsilon _{i}$ and $\beta _{i}$, which
are employed here as seed parameters. The coupling constant $\lambda _{i}$
is provided by its optimal dependence (\ref{optimal}):
\begin{equation}
\lambda _{i}=\lambda _{\Delta }\left( \kappa _{i}^{\ast }\right)
\end{equation}%
by using an estimation $\kappa _{i}^{\ast }$\ of the curvature at the
intersection point $\varepsilon _{i}$. The seed values $\left( \varepsilon
_{i}^{\ast },\beta _{i}^{\ast },\kappa _{i}^{\ast }\right) $ can be obtained
from the previous estimated values $\left( \varepsilon _{i},\beta
_{i},\kappa _{i}\right) $ by using the power expansions:%
\begin{align}
\varepsilon _{i+1}^{\ast }& =\varepsilon _{i}+\Delta e  \label{a1} \\
\beta _{i+1}^{\ast }& =\beta _{i}-\kappa _{i}\Delta e+\frac{1}{2}\zeta
_{i}^{3}\Delta e^{2},  \label{a2} \\
\kappa _{i+1}^{\ast }& =\kappa _{i}-\zeta _{i}^{3}\Delta e,  \label{a3}
\end{align}%
where $\Delta e$ is the energy step. Here, it is convenient to
consider the power-expansion up to the third derivative of the
entropy per particle $ \zeta _{i}^{3}$, since the calculation of the
fourth derivative $\zeta _{i}^{4}$ can be only obtained with a
sufficient precision after performing a very large MC run.

\subsection{Efficiency}

The efficiency of MC methods is commonly characterized in terms of the
so-called \textit{decorrelation time} $\tau $, that is, the minimum number
of MC steps needed to generate effectively independent, identically
distributed samples in the Markov chain \cite{mc3}. Its calculation in this
approach can be performed by using the expression:
\begin{equation}
\tau =\lim_{M\rightarrow \infty }\tau _{M}=\lim_{M\rightarrow \infty }\frac{%
M\cdot var\left( \varepsilon _{M}\right) }{var\left( \varepsilon _{1}\right)
},  \label{decorrelation}
\end{equation}%
where $var\left( \varepsilon _{M}\right) =\left\langle \varepsilon
_{M}^{2}\right\rangle -\left\langle \varepsilon _{M}\right\rangle ^{2}$ is
the variance of $\varepsilon _{M}$, which is defined as the arithmetic mean
of the energy per particle $\varepsilon $ over $M$ samples (consecutive MC
steps):%
\begin{equation}
\varepsilon _{M}=\frac{1}{M}\sum_{i=1}^{M}\varepsilon _{i}.
\end{equation}%
However, the decorrelation time $\tau $ only provides a partial view of the
efficiency in the case of the extended canonical MC methods discussed in
this work. In general, the efficiency is more appropriately characterized by
the number of MC steps $S$ needed to achieve the convergence of a given run.
The question is that the number of MC steps $S$ needed to achieve a
convergence of the expectation value $\left\langle x\right\rangle $ of a
given observable $x$ also depends on its thermal dispersion $\triangle x$.
For example, to obtain an estimation of the expectation value $\left\langle
x\right\rangle $ with a statistical error $\epsilon _{x}<a$, the number of
MC steps $S$ should obey the following inequality:
\begin{equation}
S>\frac{\tau \left( \Delta x\right) ^{2}}{a^{2}}  \label{S.gen}
\end{equation}%
While the fluctuating behavior of a given observable $x$ is an
intrinsic system feature in canonical MC methods, this is not longer
valid in the present framework. In this case, the fluctuating
behavior crucially depends on the nature of the external influence
acting on the system, e.g.: the thermal dispersion of the system
energy $\triangle E$ depends on the coupling constant $\lambda $ in
Eq.(\ref{dispersions}). In particular, the number of MC steps $S$
needed to obtain a point of the caloric curve $\left( \varepsilon
,\beta \right) $ with a precision $\sqrt{\epsilon _{\varepsilon
}^{2}+\epsilon _{\beta }^{2}}<a$ should be evaluated in terms of the
total dispersion $\Delta _{T}^{2}$ introduced in
Eq.(\ref{min.DeltaT}) as follows:
\begin{equation}
S=\frac{\tau \Delta _{T}^{2}}{Na^{2}}.
\end{equation}%
We refer to the quantity $\eta =\tau \Delta _{T}^{2}$ as the \textit{%
efficiency factor}. Clearly, an extended canonical MC method is more
efficient as smaller is its efficiency factor $\eta $.

\section{Applications}

\subsection{Potts model and its extended canonical MC algorithms}

For convenience, let us reconsider again the model system studied in our
previous paper \cite{vel-emc}: the $q$-state Potts model \cite{mc3}:
\begin{equation}
H_{q}=\sum_{\left\{ ij\right\} }\left( 1-\delta _{\sigma _{i}\sigma
_{j}}\right) ,  \label{Potts}
\end{equation}%
defined on the square lattice $L\times L$ with periodic boundary conditions.
The sum $\left\{ ij\right\} $ is over nearest neighbor sites, with $\sigma
_{i}=1,2,\ldots q$ being the spin variable on the $i$-th site. This model
undergoes a continuous phase transition when $q=2-4$, which turns
discontinuous for $q>4$.

The direct way to implement an extended canonical MC simulation of this
model is by using the \textit{Metropolis importance sample} \cite%
{metro,Hastings}, whose transition probability is given by:
\begin{equation}
W\left( X_{i}\rightarrow X_{j};\beta _{\omega }^{t}\right) =\min \left\{
1,\exp \left( -\beta _{\omega }^{t}\delta E_{if}\right) \right\} .
\end{equation}%
Besides this last local algorithm, the MC study of the model
(\ref{Potts}) can be also carried out by using non-local MC methods
such as the known Swendsen-Wang \cite{SW,pottsm} or Wolff's
\cite{wolf}\ clusters algorithms. As discussed elsewhere \cite{mc3},
such clusters algorithms are based on
the consideration of the \textit{Fortuin--Kasteleyn theorem} \cite{FK,FK2}:%
\begin{equation}
Z=\sum_{spins}e^{-\beta _{B}H_{q}}=\sum_{bonds}p^{b}\left( 1-p\right)
^{N_{b}-b}q^{N_{c}},
\end{equation}%
which possibilities a mapping of this model system to a random clusters
model of percolation. Here, $p=1-e^{-\beta _{B}}$ is the acceptance
probability of bonds, $N_{c}$ is the number of clusters, $b$ is the number
of bonds, and $N_{b}$ is the total number of possible bonds.

In order to implement the extended versions of these clusters algorithms,
let us denote by $p_{i}=1-e^{-\beta _{\omega }^{i}}$ the acceptance
probability of bonds by starting from the initial configuration $X_{i}$. The
transition probability of the direct process $W_{i\rightarrow f}$ can be
expressed as follows:%
\begin{equation}
W_{i\rightarrow f}=p_{i}^{b_{a}}\left( 1-p_{i}\right) ^{b_{p}+b_{d}}.
\end{equation}%
where $b_{a}$ is the number of inspected bonds which have been accepted,
while $b_{p}+b_{d}$ is the number of inspected bonds which have been
rejected. At this point, it is also important to identify the number of
rejected bonds $b_{d}$ which have been destroyed in the final configuration $%
X_{f}$, as well as the number $b_{c}$ of created bonds. Note that these
bonds are responsible of the energy varying $\delta E_{if}$ after the
transition, $\delta E_{if}=b_{d}-b_{c}$. In the reverse process, the
accepted bonds $b_{a} $ of the direct process are also accepted with
probability $p_{f}=1-e^{-\beta _{\omega }^{f}}$, while the created bonds $%
b_{c}$ as well as the rejected bonds $b_{p}$ that have not been destroyed in
the final configuration $X_{f}$ are now rejected with probability $1-p_{j}$.
Thus, the transition probability of the reverse process $W_{f\rightarrow i}$
can be expressed as follows:
\begin{equation}
W_{f\rightarrow i}=p_{f}^{b_{a}}\left( 1-p_{f}\right) ^{b_{p}+b_{c}}.
\end{equation}%
Within the canonical ensemble, where $p_{i}=p_{f}=1-e^{-\beta _{B}}$, it is
easy to see that transition probability obeys the detailed balance condition:%
\begin{equation}
\frac{W_{f\rightarrow i}}{W_{i\rightarrow f}}=\exp \left( \beta _{B}\delta
E_{ij}\right) ,
\end{equation}%
For a general case with $\lambda >0$, one should introduce the \textit{a
posteriori} acceptance probability (\ref{wif}) in order to fulfil the
detailed balance, which can be expressed as follows:%
\begin{equation}\label{wif.potts}
w_{i\rightarrow f}=\left\{ 1,\exp \left( \theta _{if}\right)
\right\} ,
\end{equation}%
where the argument $\theta _{if}$ depends on the integer numbers
$\left( b_{a},b_{p},b_{c},b_{d}\right) $:
\begin{equation}
\theta _{if}=b_{a}\log \left( \frac{p_{f}}{p_{i}}\right) -\frac{1}{2N}%
\lambda \left( b_{d}-b_{c}\right) \left( 2b_{p}+b_{c}+b_{d}\right) .
\end{equation}

\subsection{Results and discussions}

\begin{figure}[t]
\begin{center}
\includegraphics[
height=3in, width=4in ]{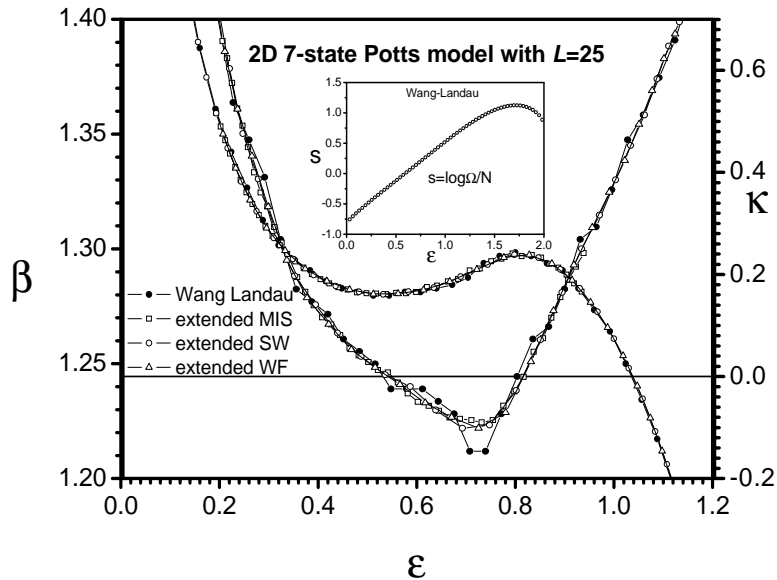}
\end{center}
\caption{Microcanonical caloric and curvature curves associated with
the $2D$ seven-state Potts model obtained from the application of
the extended canonical MC algorithms (\textbf{MIS}: Metropolis
importance sample, \textbf{SW}: Swendsen-Wang, and \textbf{WF}:
Wolff's clusters algorithm). Such results can be compared with the
ones obtained from a direct numerical differentiation of the entropy
per particle $s=\log \Omega /N$ obtained by using the Wang-Landau
sampling method.} \label{finite.eps}
\end{figure}

The results derived from the extended versions of the Metropolis
importance sample, as well as the Swendsen-Wang and Wolff's clusters
algorithms are shown in FIG.\ref{finite.eps} for the particular case
of the $2D$ seven-state Potts model with $L=25$. Each point of these
dependencies have been obtained from MC runs with $10^{6}$ steps.
For comparison purposes, we have also carried out the calculation of
the caloric $\beta \left( \varepsilon \right) $ and the curvature
$\kappa \left( \varepsilon \right) $ curves by performing a direct
numerical differentiation of the entropy per particles $
s(\varepsilon)=\log $ $\Omega (E) /N$ obtained from the Wang-Landau
sampling method \cite{WangLandau}, which is shown in the inset panel
($N=L^{2}$). Although there exist a good agreement among all these
MC results, the ones obtained from the Wang-Landau method seem to be
less significant, overall, the results corresponding to the
curvature curve $\kappa(\varepsilon)$.

\begin{figure}[t]
\begin{center}
\includegraphics[
height=3in, width=4in ]{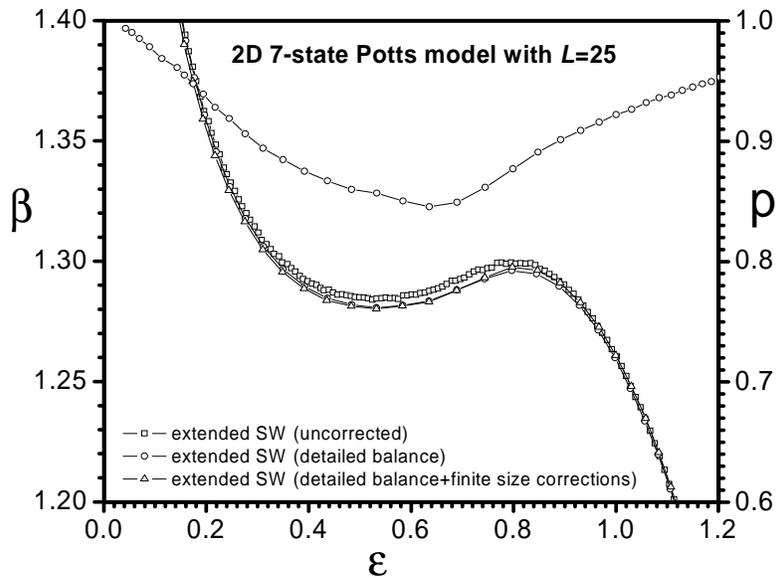}
\end{center}
\caption{Microcanonical caloric curves $\protect\beta \left( \protect%
\varepsilon \right) $\ showing the different corrections of finite size
effects corresponding to the Swendsen-Wang clusters algorithm. We also show
here the expectation value of the acceptance probability $p=\left\langle
w_{i\rightarrow f}\right\rangle $ \textit{versus} energy per particle $%
\protect\varepsilon $ corresponding to the extended Swendsen-Wang algorithm
obeying detailed balance.}
\label{corrections.eps}
\end{figure}

The previous results constitute a clear illustration that the finite size
corrections formulae introduced in subsection \ref{finite} provide a
significant improvement in the precision of this kind of MC calculations. In
this particular example with $L=25$, the contribution of first-order
correction of finite size effects in the caloric curve $\beta \left(
\varepsilon \right) $ have a typical order of $\delta _{1}\sim 10^{-3}$,
while the one corresponding to the second-order correction are of order $%
\delta _{2}\sim 10^{-5}$. Nevertheless, the most significative
correction of finite size effects in non-local canonical MC methods
comes from the fulfilment of the detailed balance condition
(\ref{detailed.balance}) obtained after the introduction of
\textit{a posteriori} acceptance probability (\ref{wif}), whose
correction has a typical order of $\delta _{db}\sim 10^{-2}$. This
fact can be appreciated in FIG.\ref {corrections.eps} for the case
of the Swendsen-Wang clusters algorithm.

\begin{figure}[t]
\begin{center}
\includegraphics[
height=3.42in, width=4in ]{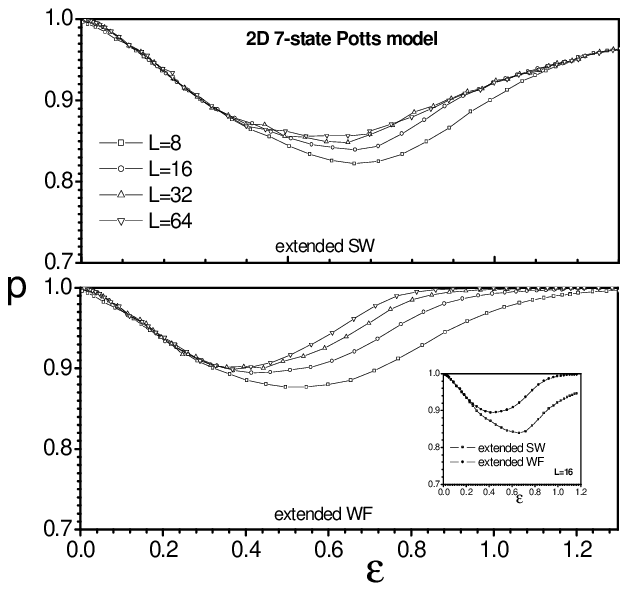}
\end{center}
\caption{Size dependence of the expectation value of the \textit{a posteriori%
} the acceptance probability $p=\left\langle w_{i\rightarrow f}\right\rangle
$ for the extended clusters algorithms.}
\label{SizeAccept.eps}
\end{figure}

Although the acceptance probability $w_{i\rightarrow f}$ for
clusters flipping is lower than the unity, its expectation value
$p=\left\langle w_{i\rightarrow f}\right\rangle $ is significantly
high for most of energy region (see also in
FIG.\ref{corrections.eps}). Since the consideration of such an
\textit{a posteriori} acceptance probability $w_{i\rightarrow f}$
corrects a finite size effect error $\delta _{\beta }=\left\vert
\beta _{\omega }^{i}-\beta _{\omega }^{t}\right\vert $ associated
with the estimation of the transition inverse temperature $\beta
_{\omega }^{t}$, one should observe a growth of the expectation
value $\left\langle w_{i\rightarrow f}\right\rangle $ as $N$
increases. Such a behavior is indeed appreciated in
FIG.\ref{SizeAccept.eps} for the extended clusters algorithms. While
all these dependencies have a similar qualitative behavior, the
expectation value of the acceptance probability $\left\langle
w_{i\rightarrow f}\right\rangle $ of the extended Wolff's clusters
algorithm in the paramagnetic region is larger than the one
corresponding to the extended Swendsen-Wang method. This is a quite
expected qualitative result. The acceptance probability
$w_{i\rightarrow f}$ should correct in the first case the finite
size error $\delta _{\beta }=\left\vert \beta _{\omega }^{i}-\beta
_{\omega }^{t}\right\vert$ related to the flipping of only one
cluster, while the finite size error $\delta _{\beta}$ related to
the extended Swendsen-Wang method is larger due to this algorithm
involves the flipping of all system clusters. At low energies or in
the ferromagnetic region, these methods have practically the same
performance, since the number of spins belonging to the cluster is
comparable to the system size $N$
(see a more detailed comparison in the inset panel of FIG.\ref%
{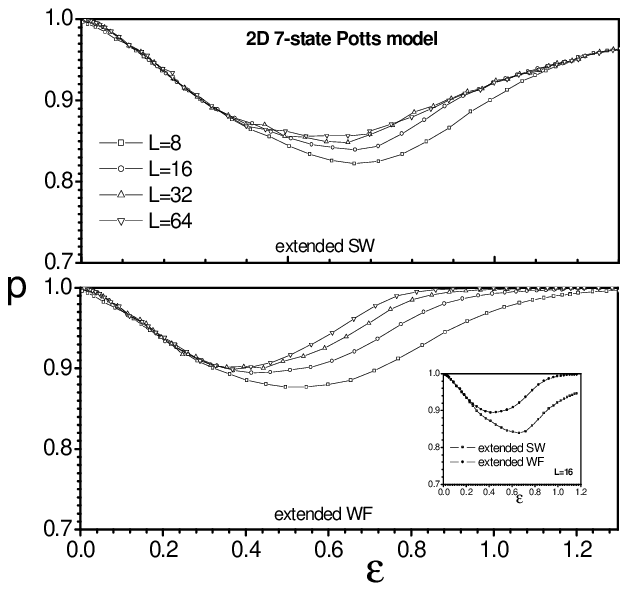}).

The qualitative behavior of the dependence $\left\langle w_{i\rightarrow
f}\right\rangle $ \textit{versus} $\varepsilon $ finds a simple explanation
in terms of the explicit dependence of the acceptance probability $%
w_{i\rightarrow f}$ on the coupling constant $\lambda $. While the
acceptance probability $w_{i\rightarrow f}\equiv 1$ within the
canonical ensemble where $\lambda =0$, this quantity undergoes a
reduction with the increase of the coupling constant $\lambda $.
Moreover, the optimal value of coupling constant $\lambda _{\Delta
}=\sqrt{1+\kappa ^{2}}-\kappa $ employed in the present MC
simulations increases with the reduction of the system curvature
$\kappa $. Indeed, the lowest values of $\left\langle
w_{i\rightarrow f}\right\rangle $ are observed in the energy region
where the system curvature also exhibits its lower values, that is,
the anomalous region with negative heat capacities $C<0$.

\begin{figure}[t]
\begin{center}
\includegraphics[
height=3in, width=4in ]{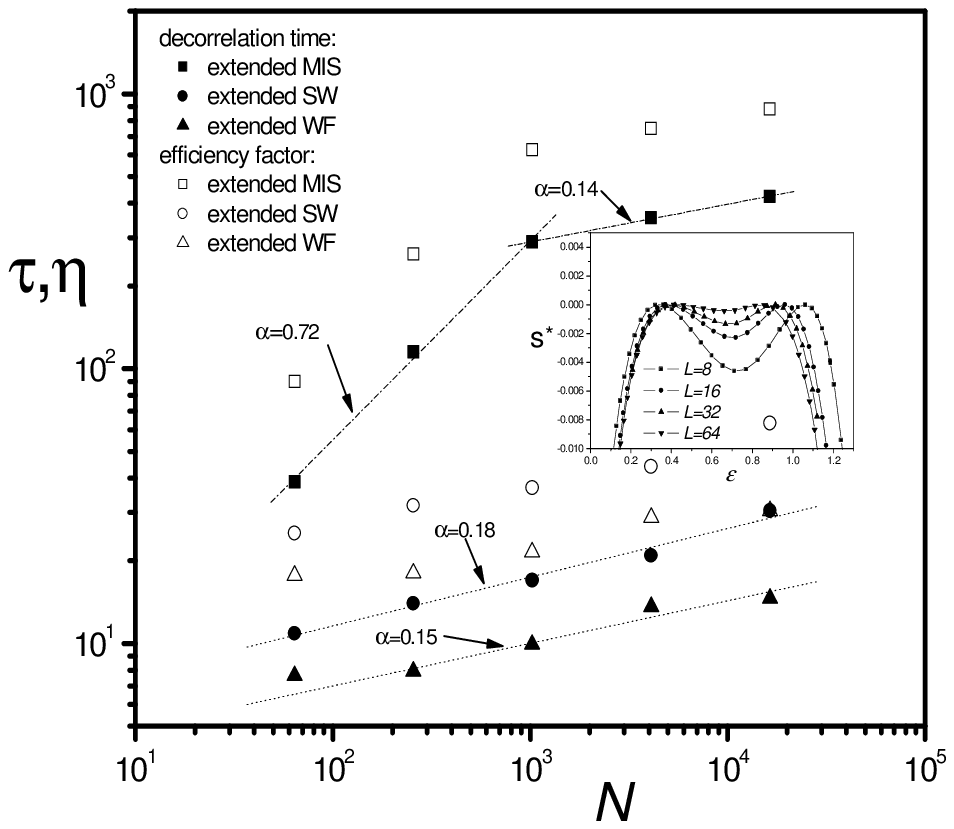}
\end{center}
\caption{Main panel: Size dependence of the decorrelation time $\protect\tau
$ and the efficiency factor $\protect\eta =\protect\tau \Delta _{T}^{2}$ for
the extended canonical MC simulations with environment inverse temperature $%
\protect\beta _{\protect\omega }(\protect\varepsilon )=\protect\beta _{e}+%
\protect\lambda (\protect\varepsilon -\protect\varepsilon _{e})$. Here, $%
\protect\beta _{e}=\protect\beta _{pt}$, $\protect\varepsilon _{e}=\protect%
\varepsilon _{2}$, $\protect\lambda =\protect\lambda _{\Delta
}\left( \protect\kappa _{2}\right) $ and $\protect\kappa
_{2}=\protect\kappa \left( \protect\varepsilon _{2}\right) $, where
$\protect\beta _{pt}$ is the estimated inverse temperature that
corresponds to the discontinuous PT, while $\protect\varepsilon
_{2}$ is the stationary solution derived from the thermal
equilibrium condition $\protect \beta (\protect\varepsilon
_{2})=\protect\beta _{pt}$ located  within the region with negative
heat capacities $C<0$. Inset panel:
Size dependence of the entropy per particle $s(\varepsilon)$, or more exactly, the quantity $%
s^{\ast }\left( \protect\varepsilon \right) =s(\protect\varepsilon )-\protect%
\beta _{pt}\protect\varepsilon +const$, which allows to show the
convex intruder associated with the region with negative heat
capacities. The energies corresponding to the two maxima $\left(
\protect\varepsilon _{1}, \protect\varepsilon _{3}\right) $ and the
minimum $\protect\varepsilon _{2}$ are the three stationary
solutions derived from the thermal equilibrium condition
$\protect\beta \left( \protect\varepsilon _{1,2,3}\right) =\protect
\beta _{pt}$. } \label{decorrelation.eps}
\end{figure}

The size dependencies of the decorrelation time $\tau $ and the
efficiency factor $\eta =\tau \Delta _{T}^{2}$ within the anomalous
region with $C<0$ are shown in the main panel of
FIG.\ref{decorrelation.eps}. Due to computational limitations, the
data were obtained from MC simulations with lattice sizes $L=\left(
8,16,32,64,128\right) $. Since the absolute values of the curvature
$\kappa $ are close to zero $\left\vert \kappa \right\vert \simeq
0$, the total dispersion is approximately given by the constant
value $\Delta _{T}^{2}\simeq 2$, in accordance with
Eq.(\ref{optimal}). This is the reason why the dependencies
$\tau(N)$ and $\eta(N)$ are almost displaced in a constant value
along the vertical direction of the log-log graph shown in
FIG.\ref{decorrelation.eps}.

As expected, the extended version of the Metropolis importance
sample exhibits the largest values of the decorrelation time $\tau $
for the lattice sizes $L$ studied in this work. Curiously, the size
dependence of its decorrelation time $\tau(N) $ shows an abrupt
transition from a power-law regime with exponent $\alpha _{1}\simeq
0$.$72$ to other one with exponent $\alpha _{2}\simeq 0$.$14$ close
to $N\simeq 10^{3}$ ($ L=32$). Despite of the extended Metropolis
importance sample is a local MC method, the effective exponent
$\alpha _{2}$ for the larger system sizes is comparable to the ones
associated with the extended clusters MC methods, $\alpha
_{SW}\simeq 0$.$18$ (Swendsen-Wang) and $\alpha _{WF}\simeq 0$.$15$
(Wolff). Note also that the efficiency of these extended clusters
methods are still significant despite of the consideration of the
\textit{a posteriori} acceptance probability (\ref{wif.potts})
employed here to ensure the detailed balance condition
(\ref{detailed.balance}).

The above examples confirm that the efficiency achieved with the
application of the present methodology is more significant than the
improvement considered by the application of re-weighting techniques
such as multicanonical method and its variant, whose typical values
of the exponent $\alpha$ ranges from 2 to 2.5 in the case of Potts
models \cite{BergM}. As already evidenced in FIG.\ref{finite.eps},
while the Wang-Landau sampling method provides a good estimation of
the entropy per particle $s(\varepsilon)$, its underlying
statistical errors are still appreciable in the curvature curve
$\kappa \left( \varepsilon \right) =-\partial ^{2}s\left(
\varepsilon \right) /\partial \varepsilon ^{2}$ regardless the
simulation was extended until the modifying factor fulfills the
condition $f_{i}<\exp \left( 10^{-10}\right) $. Such an observation
evidences that the results obtained from this last method are not
sufficiently relaxed to provide a precise estimation of the
curvature curve $\kappa \left( \varepsilon \right)$. By considering
the CPU time-cost needed to achieve the convergence, it is more
convenient to perform a punctual estimation of the caloric $ \beta
\left( \varepsilon \right) $ and curvature $\kappa \left(
\varepsilon \right) $ curves with any extended canonical MC method
instead of carrying out a numerical differentiation of the
microcanonical entropy $s(\varepsilon)$ over an energy range
obtained from a re-weighting technics such as the Wang-Landau
sampling method.

\section{Conclusions}

In this work, we have shown that the methodology to extend canonical
MC methods inspired on the consideration of the recently obtained
fluctuation relation (\ref{fdr}) can be improved to account for the
existence of finite size effects and to fulfil of the detailed
balance condition (\ref{detailed.balance}). Remarkably, despite of
the consideration of the \textit{a posteriori} acceptance
probability (\ref{wif}) reduces the efficiency of the clusters MC
methods, it has been shown that the relaxation times needed to
ensure the convergence are more significant than the ones achieved
with re-weighting technics such as multicanonical methods and its
variants. For the particular case of the seven-states Potts model,
the consideration of any extended canonical MC algorithm enables a
suppression of the super-critical slowing down associated with the
occurrence of the temperature driven first-order phase transition of
this model from an exponential growth to a very weak power-law
dependence with exponent $\alpha=0.14-0.18$.

There is still some open questions in regard to the potentialities
of the present methodology. For example, although this method has
been specially conceived to overcome the slow relaxations of
canonical MC simulations near to a region with a first-order phase
transition, in principle, there is no limitation that the same one
can be also employed to improve canonical MC simulations near to a
critical point of a continuous phase transition. Moreover, a similar
extension is possible to carry out for those MC methods based on the
consideration of the Boltzmann-Gibbs distributions:
\begin{equation}\label{BG}
dp_{BG}(E,X)=\frac{1}{Z(\beta,Y)}\exp\left[-\beta(E+YX)\right]dEdX
\end{equation}
to account for the existence of anomalous values in other response
functions besides the heat capacity. The analysis of these questions
deserve a special attention in future works.

\end{document}